# Importance of the time acquisition difference between DMSP/OLS and SNPP/VIIRS/DNB and the RAW trends in Europe

Alejandro Sánchez de Miguel[1,2,3] and Sara Krupansky[4]

[1]*Environment and Sustainability Institute, University of Exeter, Penryn, Cornwall TR10 9FE, U.K.*
[2]*Depto. de Física de la Tierra y Astrofísica, Fac. CC. Físicas, Universidad Complutense de Madrid, Plaza de las Ciencias 1, E-28040, Spain*
[3]*Instituto de Física de Partículas y del Cosmos, IPARCOS, Madrid, Spain*
[4]*SaveStars Consulting S.L. , C/Palencia 1. Las Rozas de Madrid (28231). España*

*Corresponding Author: A. Sánchez de Miguel (alejasan@ucm.es)

**Abstract**

Using the SNPP-VIIRS/DNB and the DMSP-OLS it is possible to have an idea of the evolution of the light pollution until the LEDs started to appear massively. Another of the issues of dealing with these two datasets is the different time of acquisition. In general this means that countries get dimmer late at night, although that is not always true. A counterexample of this is the country of the Netherlands because of the opening of the shutters late at night. Other countries have a much bigger drop late at night because of some practices like the total or partial turn off of public and private lighting but also dimming policies. Examples of this are countries like France, Austria and the UK.

1. Introduction

The only currently available data globally available are the data from the Defense Meteorological Satellite Program (DMSP) constellation, from 1992 until 2014 and the Suomi-North Polar Partnership(SNPP) and it's instrument Visible Infrared Imaging Radiometer Suite(VIIRS) on the Day Night Band (DNB), for the period 2012 until the present day.

The DMSP-OLS was not designed to perform night time radiometry, as was only designed to detect clouds at night, so since 1997 until today there have been numerous efforts to calibrate and inter-calibrate these satellites. Both satellites have similar spectral response on the consideration that both have panchromatic sensors although they still have significant differences. These differences, make that can be a difference on measured radiance up to 10% because of the different spectra of the sources. Also have a different spatial resolution, as the DMSP - OLS has a PSF (Point Spread Function) of 5 km and the VIIRS of 750 m (so no confuse this with the spatial sampling, that is 30 arc second and 15 arc seconds), but is always possible to degrade the resolution of the VIIRS image to the DMSP or collect data on a regional basis big enough to collect most of the signal from one country or region. But, probably, the most important difference is the time of the acquisition. This variable can be more relevant or less depending on several factors. We should remember that the DMSP-OLS used to acquire the data around 21:30 solar time, and the VIIRS around 1:30 solar time.

As an example, we are going to show the comparison between Belgium, France and the Netherlands. These three countries situated on the north of Europe, have a very different behavior when we compare the "RAW" data (aka. without using offsets to intercalibrate VIIRS and DMSP). The source of this data is [0]. On the Figs. A1, A2 and A3 the green dots correspond to the DMSP calibrated, blue to DMSP non calibrated and red to VIIRS. On Fig A1. we can see that there is not a big difference between the blue dots and the green ones. This indicates that the saturated areas are not too relevant, and most of the changes are in general happening on the city edges. Similar happens on the Fig. A2, A3 and even A1, which from these three countries has the most important gap (~11%). Then, we can see a smaller gap between the blue dots and red dots of the VIIRS. This means that the amount of ornamental lights and dim public street lighting is not very relevant (~8%). But, when we compare with the other countries, we see a much bigger difference. In the case of France, the drop is nearly 50% and in the case of the Netherlands an increase of 83%. The first thing that we should notice is that each country goes in a different direction. In general, all countries of the EU keep the same trend (see Belgium Fig. A1, Spain Fig. A5, Greece Fig. A7, United Kingdom Fig. A6, Portugal Fig. A9, Ireland Fig. A10 ) or have a big drop (see France Fig. A2, Germany Fig. A4, Italy Fig. A8, Czech Republic Fig. A11, Poland Fig. A12, Croatia Fig. A13, Denmark Fig. A14, Finland Fig. A15, Sweden Fig. A16, Slovakia Fig. A17, Romania Fig. A18). So, the case of the Netherlands is unique in this region of the world and it is explained by the intensive use of green houses and its combination with the laws of use of shutters on these green houses.

Even though the comparison of the satellites is based on "RAW" data, graphs of sunset and midnight time emissions have been added for readers' comparison.

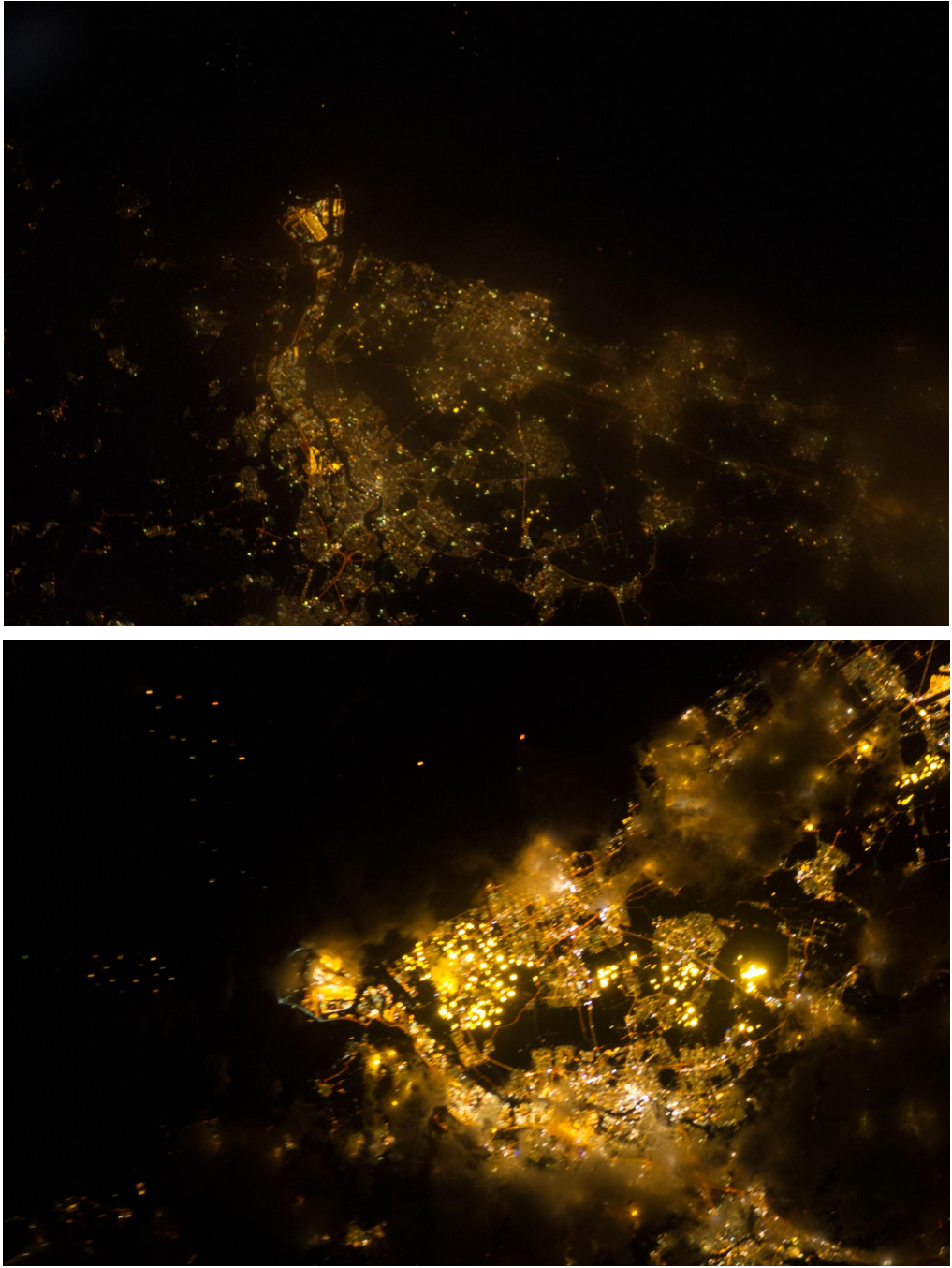

*Fig. D1*: Two images of the area of Rotterdam. Taken both with a Nikon D3S and a 180 mm lens. Clearly it can be seen that even if both images are taken at the same time of the year, the time of acquisition is very different. The first one, was taken close to the DMSP typical acquisition time, and second close to the VIIRS acquisition time. There is a significant difference between them and is clearly due to the yellow green houses that are very bright. ISS030-E-296115, 2012.02.20, 19:29:29 GMT, ISS034-E-39336, 2013.02.02 23:40:50 GMT.

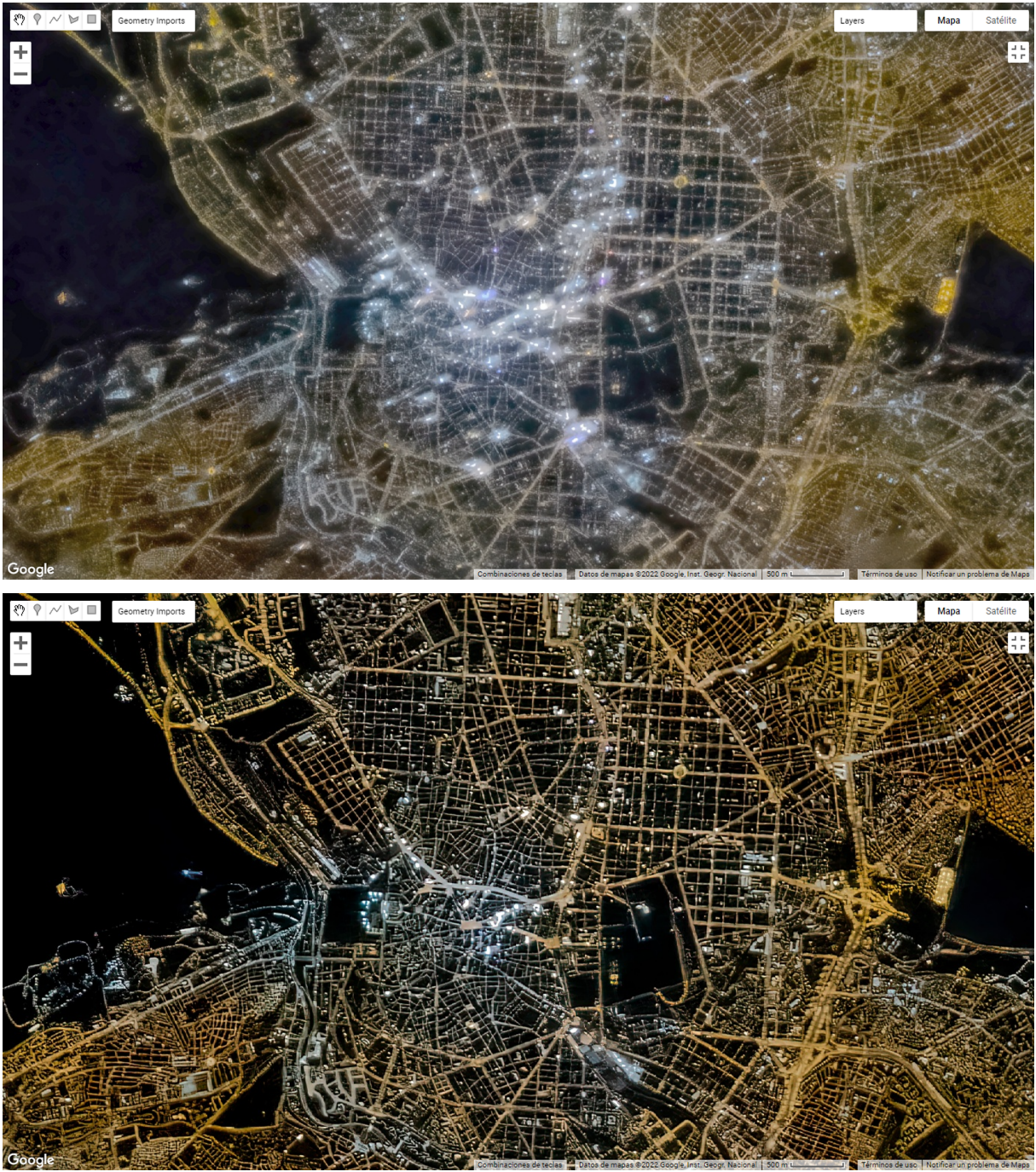

*Fig. D2*: Two images of the area of Madrid. Taken both with a Nikon D5 and a 400 mm lens. Clearly can be seen off the top image has a larger number of hotspots. First image was taken during the first part of the night and the second near midnight. ISS065-E-045335 2021.05.13 21:38:14 GMT, ISS065-E-203751, 2021.07.24, 23:35:05 GMT. Interactive version available on: https://pmisson.users.earthengine.app/view/madridlight.

About the big decrees on VIIRS, as mentioned, it can be due to the turn off of ornamental lights[3] or dimming of public street lighting. In this direction, France is known to have a significant amount of comunes (estimated 33%, confirmed 23%), which leads to total or partial extinction of public lighting. This policy has been evaluated on the topic of bat protection.

This practice is also common in some villages of the United Kingdom, Austria[5] and Germany and probably many other countries. Other places, like Italy have light pollution laws, like Croatia, Slovenia or Czech Republic. Spain has an energy efficiency regulation, but with exceptions. In the period of 2012-2013 there were no examples of significant dimming, apart from the ornamental lights which are mainly on the saturated parts so do not interfere with the non calibrated DSMP. The Gap between the calibrated DMSP and the non calibrated DSMP suggest that the ornamental lights could be 28% of the light[7-8]. In some cities like Madrid there are regulations to dimme the ornamental lights after a certain time. The law of 2008 in Spain also made mandatory to have a dimming system but typically the reduction was only on a 20% and not widely implemented, as we can deduce from the IDAE document, as the effective hours of work where 85% of the reference (aka. 3511 h vs 4100 h)[4]. Splain is also one of the countries with per light per capita as Portugal[7]. This is explain because of its high population density per built area[8].

In summary, neither VIIRS or DMSP calibrated can be used as absolute baseline of radiance, as the radiance emitted is a dynamic variable through the night. These differences can be more or less important deeping of the local policies. Under the current lack of information, our compromise has been assuming that the DMSP calibrated is a good base line for the beginning of the night, and the VIIRS is for the late night.

## *Appendix*

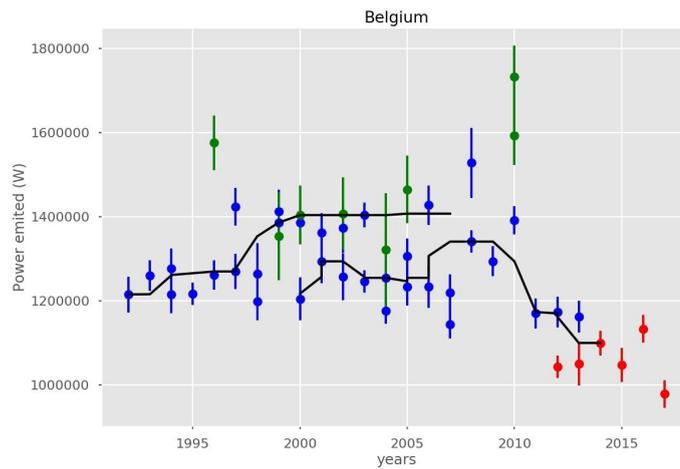

*Fig. A1*: RAW trends of Belgium. Green data correspond to the calibrated DMSP, blue correspond to non-calibrated DMSP and red correspond to VIIRS. Can be seen that there is not a big gap between the DMSP non-calibrated data and the VIIRS.

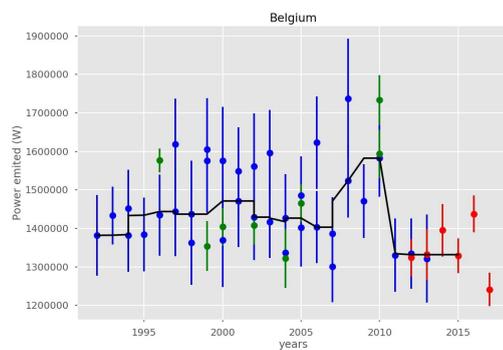 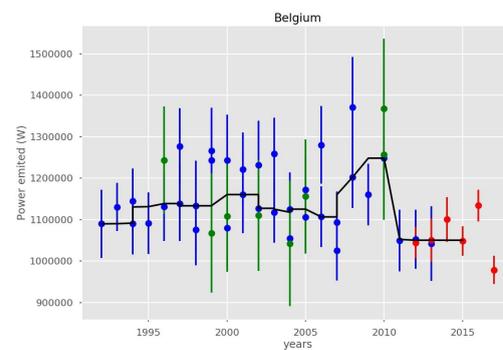

*Fig. B1*: Sunset trends of Belgium.            *Fig. C1*: Midnight trends of Belgium.

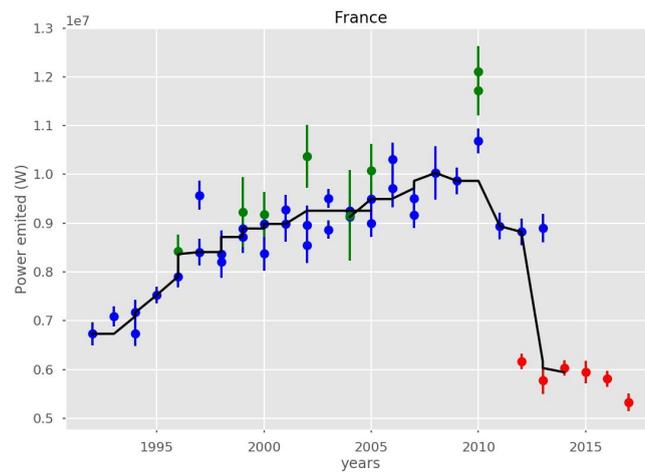

*Fig. A2*: RAW trends of France. Green data correspond to the calibrated DMSP, blue correspond to non-calibrated DMSP and red correspond to VIIRS. Can be seen that there is a big gap between the DMSP non-calibrated data and the VIIRS. In this case we explain it because of dimming of public lighting[6].

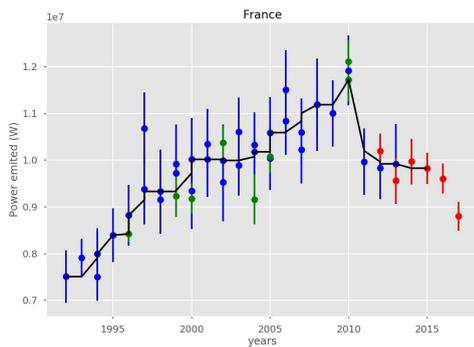
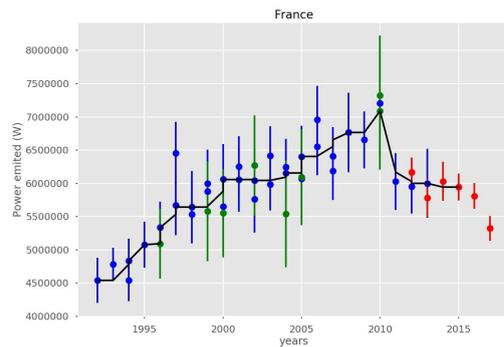

*Fig. B2:* Sunset trends of France        *Fig. C2:* Midnight trends of France

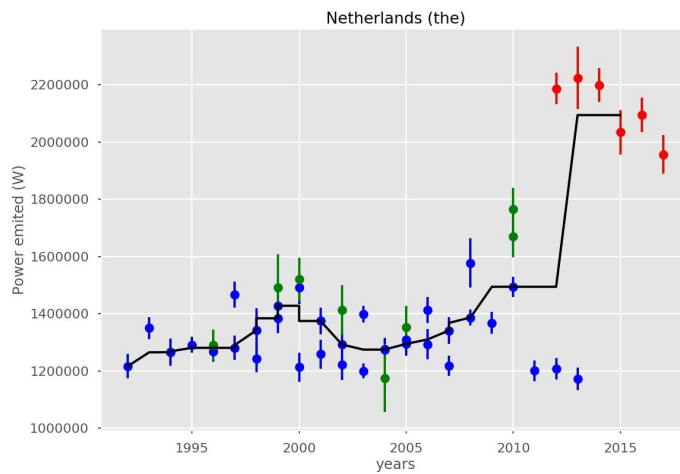

*Fig. A3*: RAW trends of the Netherlands. Green data correspond to the calibrated DMSP, blue correspond to non-calibrated DMSP and red correspond to VIIRS. Can be seen that there is a big gap between the DMSP non-calibrated data and the VIIRS. In this case we explain it because of the greenhouse policy with the shutters, getting brighter.

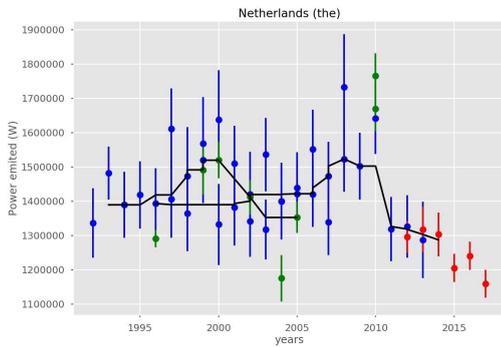 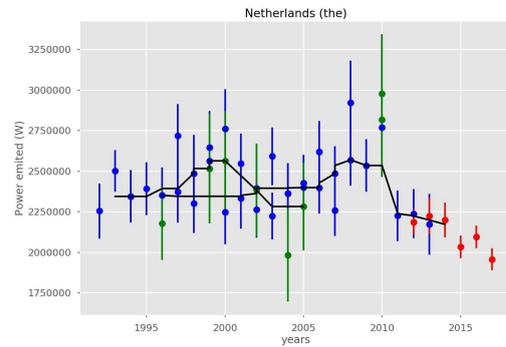

*Fig. B3:* Sunset trends of the Netherlands.   *Fig. C3:* Midnight trends of the Netherlands.

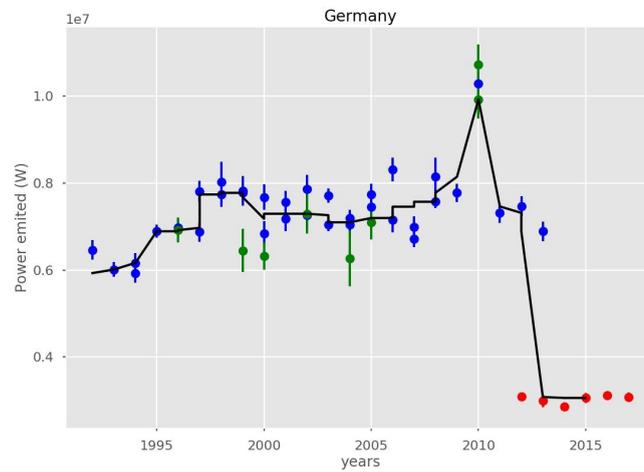

*Fig. A4*: RAW trends of Germany. Green data correspond to the calibrated DMSP, blue correspond to non-calibrated DMSP and red correspond to VIIRS. Can be seen that there is a big gap between the DMSP non-calibrated data and the VIIRS. Probably because of the dimming of public lighting.

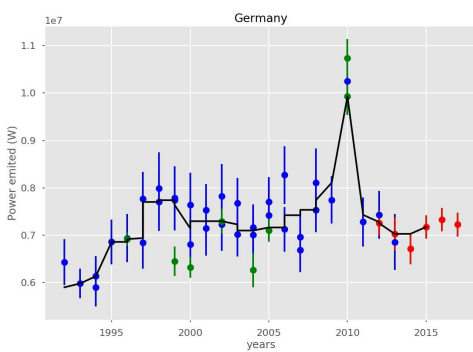 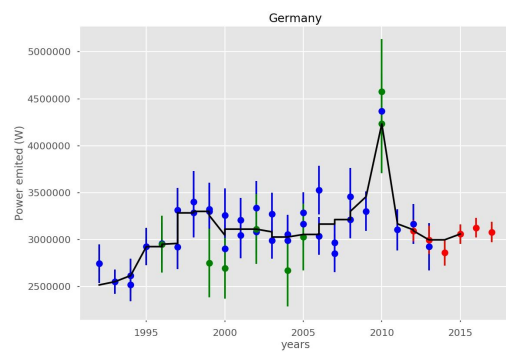

*Fig. B4:* Sunset trends of Germany.   *Fig. C4:* Midnight trends of Germany.

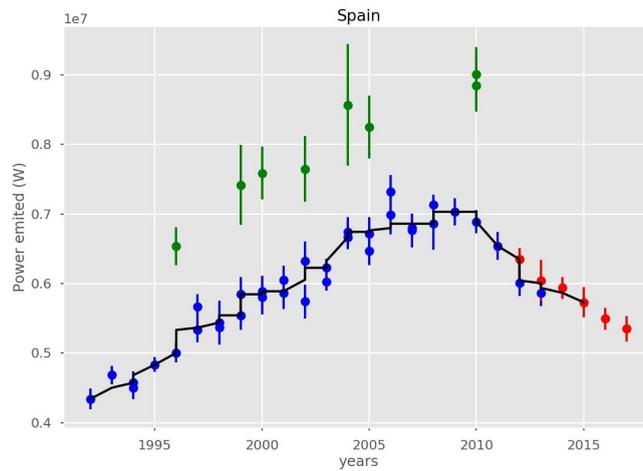

*Fig. A5*: RAW trends of Spain. Green data correspond to the calibrated DMSP, blue correspond to non-calibrated DMSP and red correspond to VIIRS. Can be seen that there is not a big gap between the DMSP non-calibrated data and the VIIRS. There is a gap between DMSP calibrated and the VIIRS, probably explained by the dimming of ornamental lights.

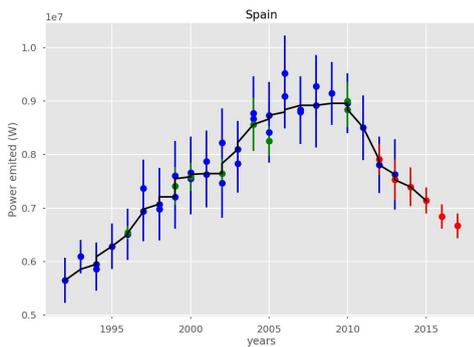
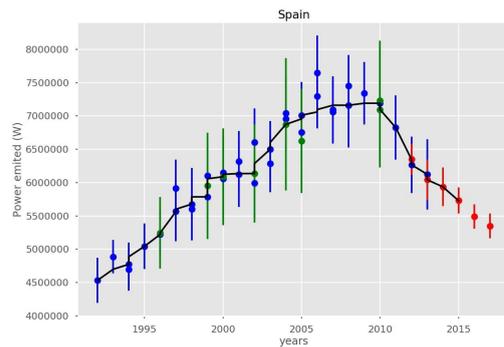

*Fig. B5:* Sunset trends of Spain.  *Fig. C5:* Midnight trends of Spain.

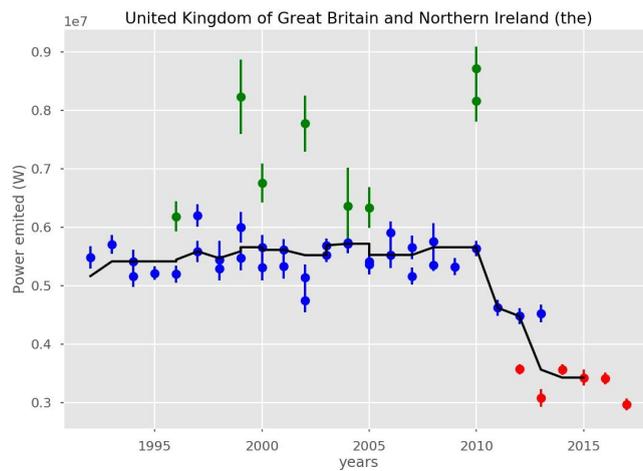

*Fig. A6*: RAW trends of the United Kingdom. Green data correspond to the calibrated DMSP, blue correspond to non-calibrated DMSP and red correspond to VIIRS. Can be seen that there

is not a big gap between the DMSP non-calibrated data and the VIIRS. Probably because of the dimming of public lighting.

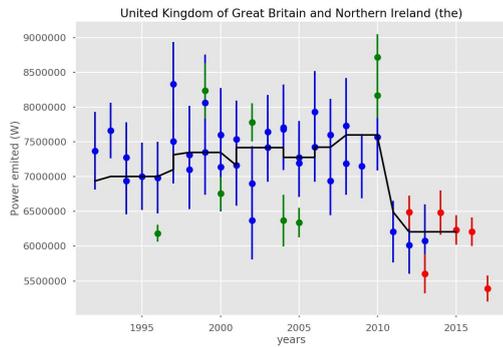

*Fig. B6:* Sunset trends of the United Kingdom.

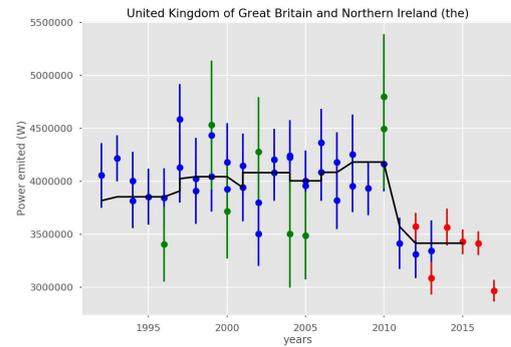

*Fig. C6:* Midnight trends of the United Kingdom.

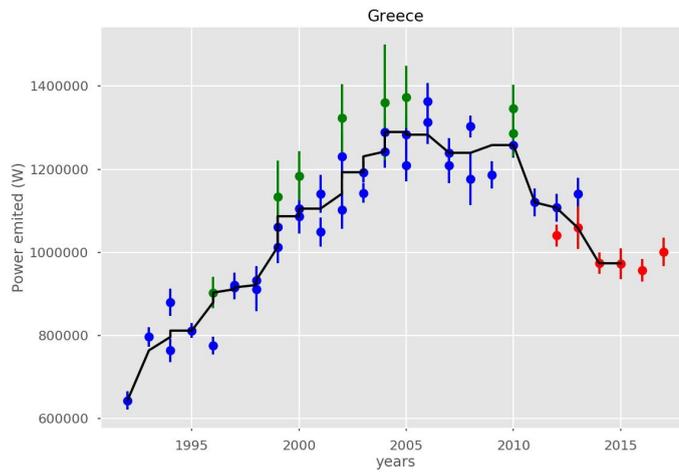

*Fig. A7:* RAW trends of Greece. Green data correspond to the calibrated DMSP, blue correspond to non-calibrated DMSP and red correspond to VIIRS. Can be seen that there is not a big gap between the DMSP non-calibrated data and the VIIRS.

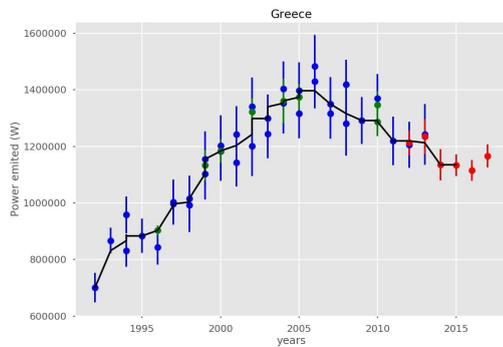

*Fig. B7:* Sunset trends of Greece.

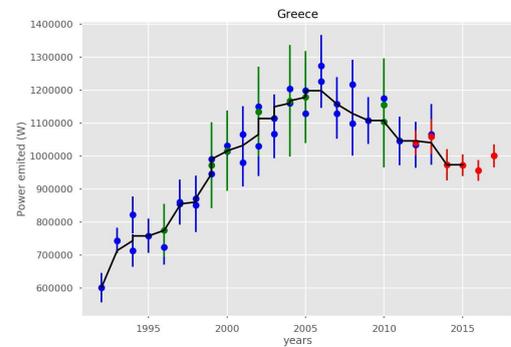

*Fig. C7:* Midnight trends of Greece.

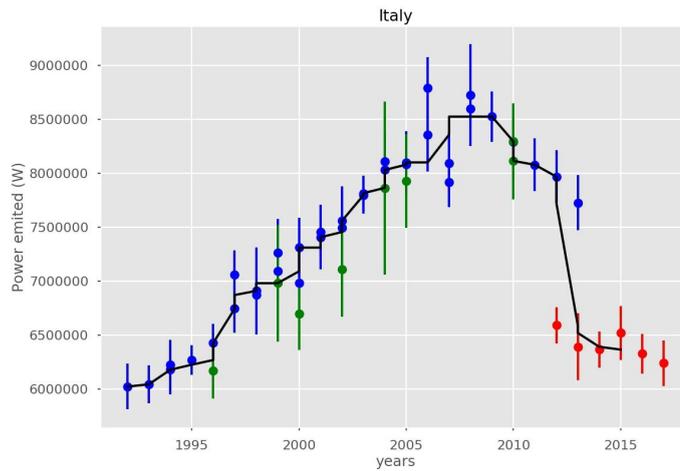

*Fig. A8*: RAW trends of Italy. Green data correspond to the calibrated DMSP, blue correspond to non-calibrated DMSP and red correspond to VIIRS. Can be seen that there is a big gap between the DMSP non-calibrated data and the VIIRS. Probably because of the dimming of public lighting.

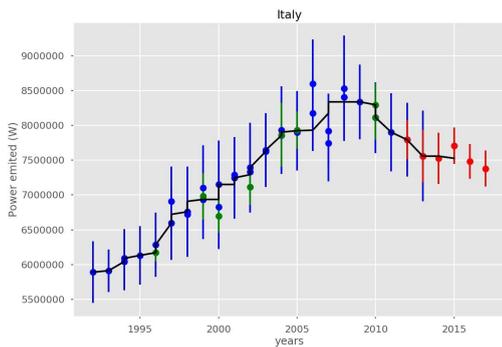 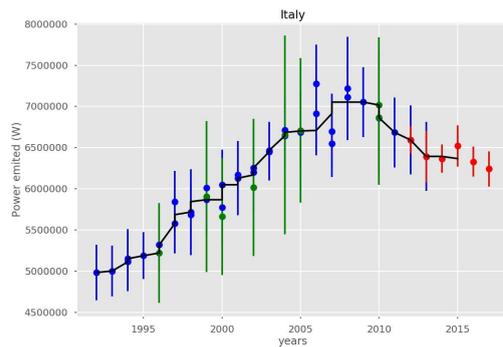

*Fig. B8:* Sunset trends of Italy.  *Fig. C8:* Midnight trends of Italy.

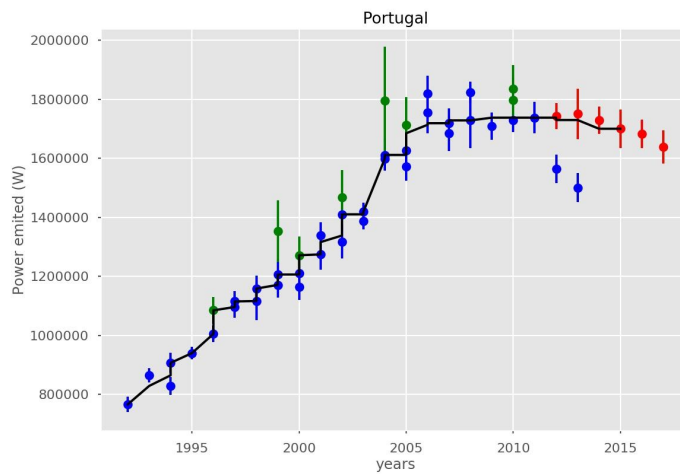

*Fig. A9*: RAW trends of Portugal. Green data correspond to the calibrated DMSP, blue correspond to non-calibrated DMSP and red correspond to VIIRS. Can be seen that there is not a big gap between the DMSP non-calibrated data and the VIIRS. Probably because of the dimming of public lighting.

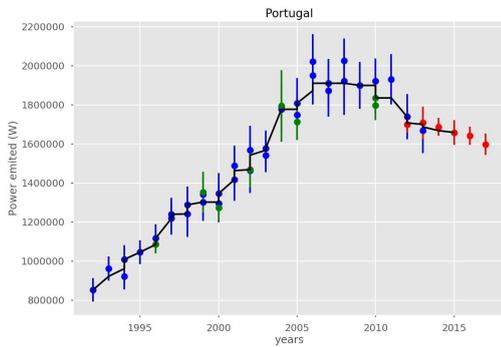

*Fig. B9:* Sunset trends of Portugal.

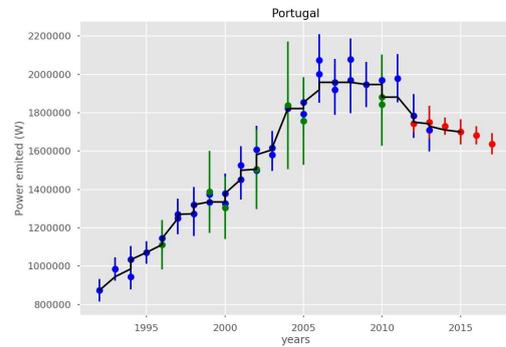

*Fig. C9:* Midnight trends of Portugal.

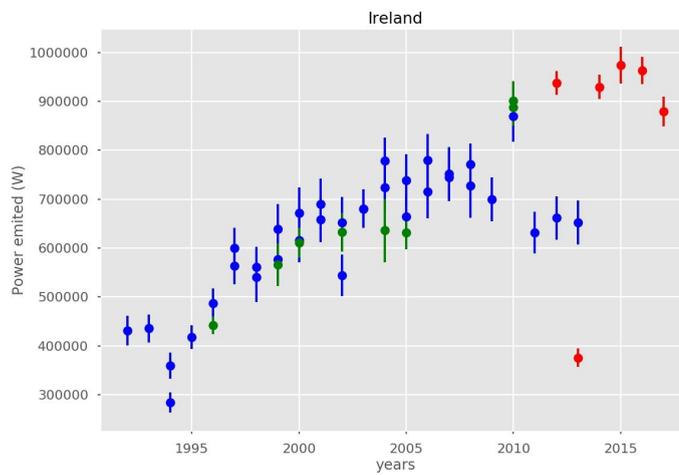

*Fig. A10:* RAW trends of Ireland. Green data correspond to the calibrated DMSP, blue correspond to non-calibrated DMSP and red correspond to VIIRS. Can be seen that there is not a big gap between the DMSP non-calibrated data and the VIIRS. Probably there is no dimming.

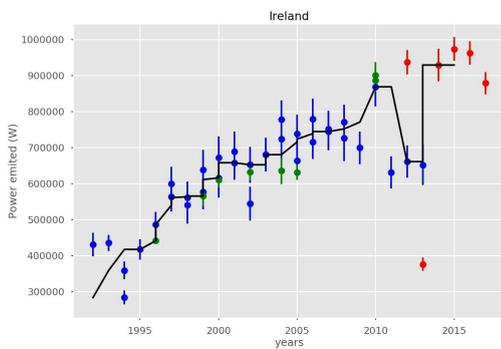

*Fig. B10:* Sunset trends of Ireland.

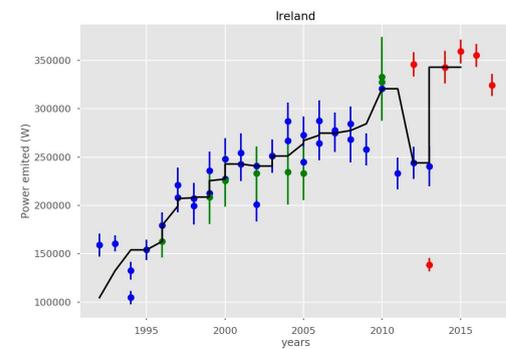

*Fig. C10:* Midnight trends of Ireland.

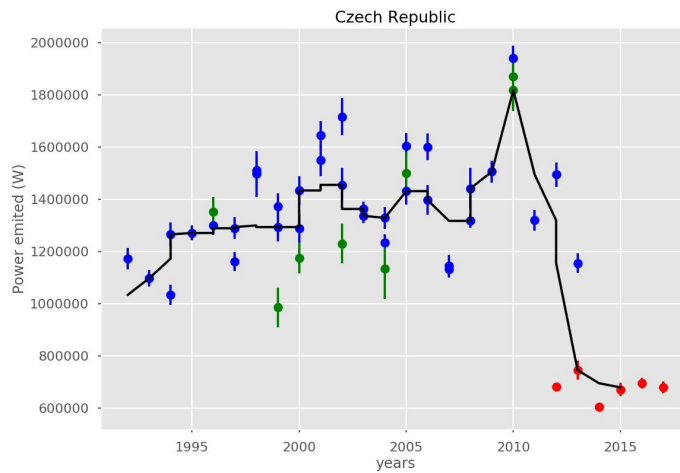

*Fig. A11*: RAW trends of Czech Republic. Green data correspond to the calibrated DMSP, blue correspond to non-calibrated DMSP and red correspond to VIIRS. Can be seen that there is a big gap between the DMSP non-calibrated data and the VIIRS. Probably because of the dimming of public lighting.

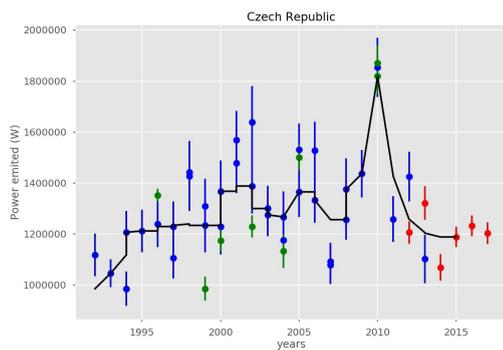

*Fig. B11:* Sunset trends of Czech Republic.

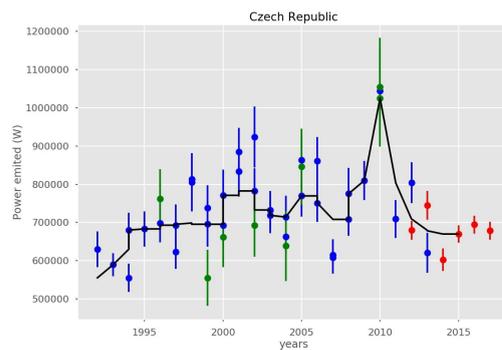

*Fig. C11:* Midnight trends of Czech Republic.

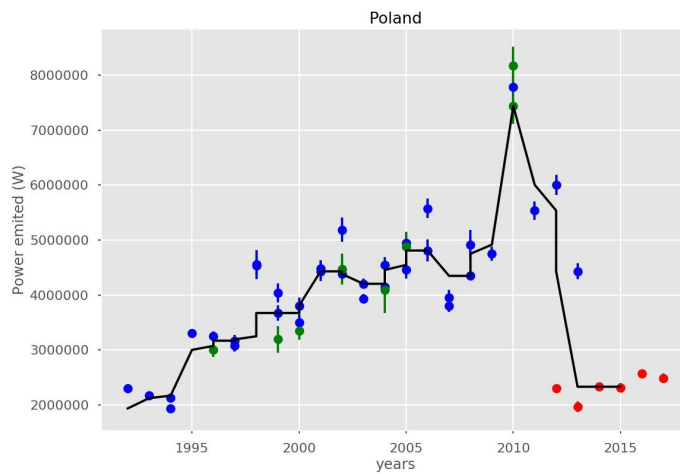

*Fig. A12*: RAW trends Poland. Green data correspond to the calibrated DMSP, blue correspond to non-calibrated DMSP and red correspond to VIIRS. Can be seen that there is a big gap between the DMSP non-calibrated data and the VIIRS. Probably because of the dimming of public lighting.

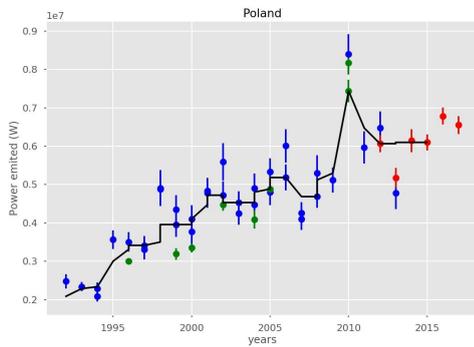 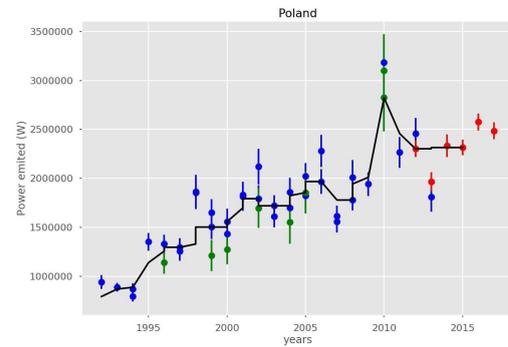

*Fig. B12:* Sunset trends of Poland    *Fig. C12:* Midnight trends of Poland.

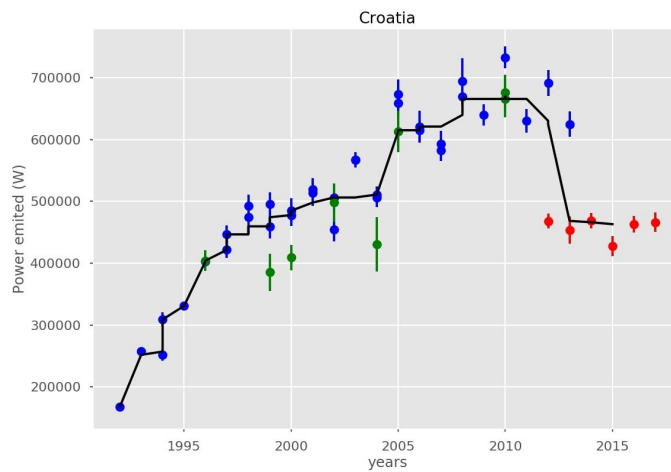

*Fig. A13*: RAW trends of Croatia. Green data correspond to the calibrated DMSP, blue correspond to non-calibrated DMSP and red correspond to VIIRS. Can be seen that there is a big gap between the DMSP non-calibrated data and the VIIRS. Probably because of the dimming of public lighting.

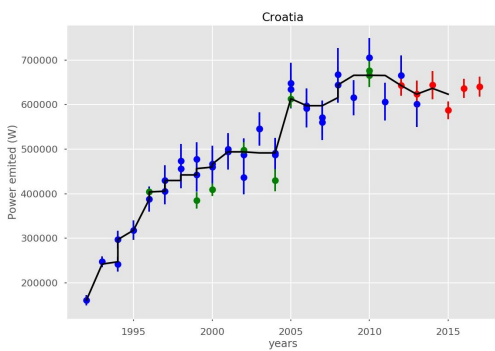 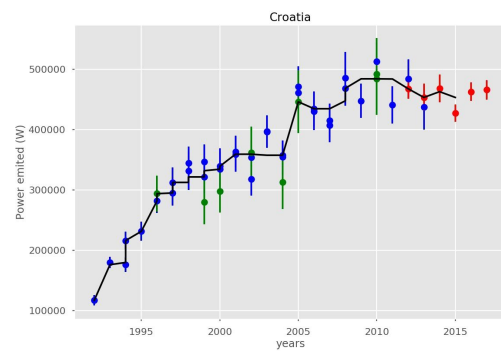

*Fig. B13:* Sunset trends of Croatia.    *Fig. C13:* Midnight trends of Croatia.

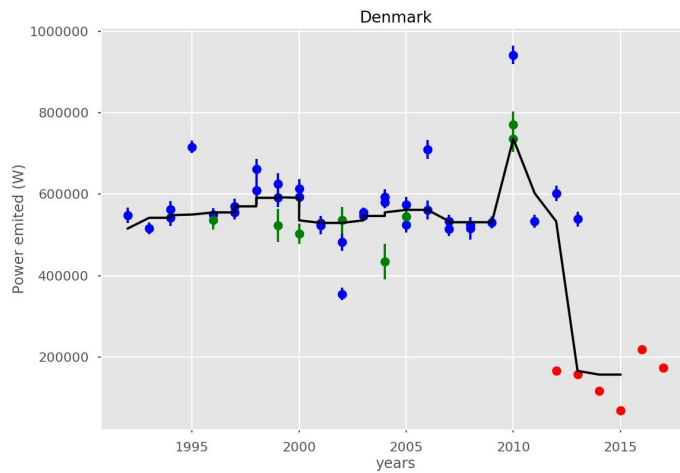

*Fig. A14*: RAW trends of Denmark. Green data correspond to the calibrated DMSP, blue correspond to non-calibrated DMSP and red correspond to VIIRS. Can be seen that there is a big gap between the DMSP non-calibrated data and the VIIRS. Probably because of the dimming of public lighting.

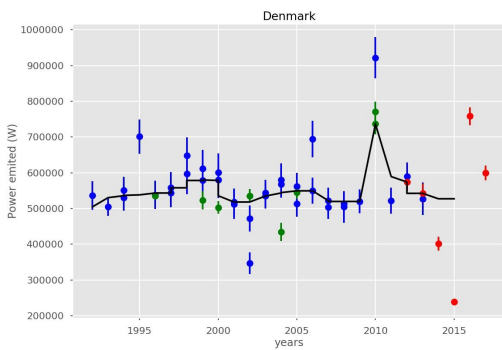

*Fig. B14:* Sunset trends of Denmark.

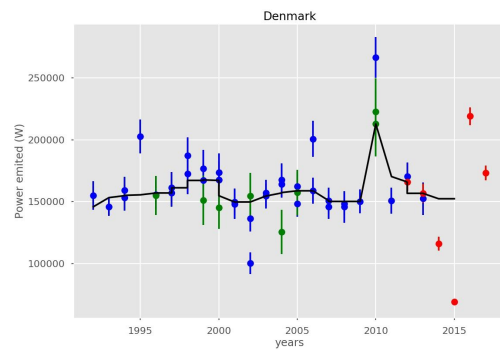

*Fig. C14:* Midnight trends of Denmark.

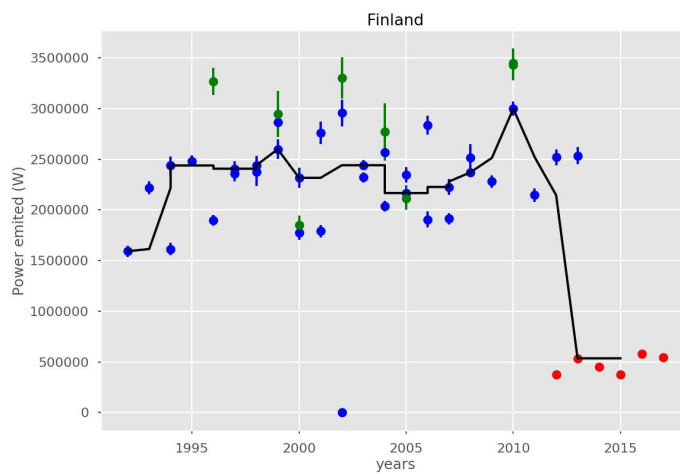

*Fig. A15*: RAW trends of Finland. Green data correspond to the calibrated DMSP, blue correspond to non-calibrated DMSP and red correspond to VIIRS. Can be seen that there is a big gap between the DMSP non-calibrated data and the VIIRS. Probably because of the dimming of public lighting.

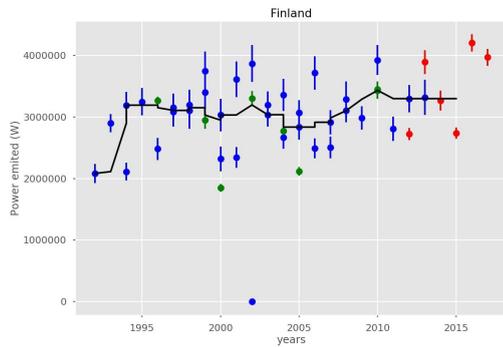 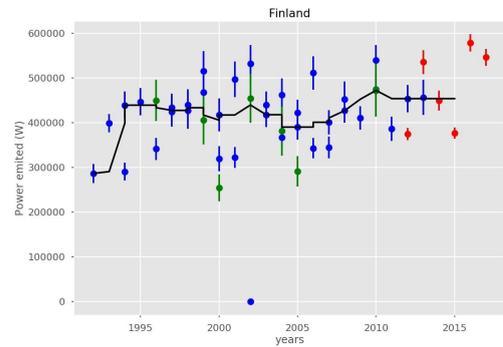

*Fig. B15:* Sunset trends of Finland.    *Fig. C15:* Midnight trends of Finland.

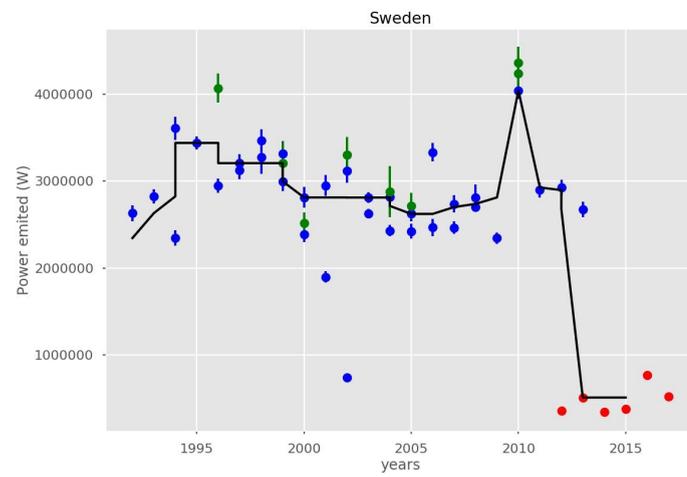

*Fig. A16*: RAW trends of Sweden. Green data correspond to the calibrated DMSP, blue correspond to non-calibrated DMSP and red correspond to VIIRS. Can be seen that there is a big gap between the DMSP non-calibrated data and the VIIRS. Probably because of the dimming of public lighting.

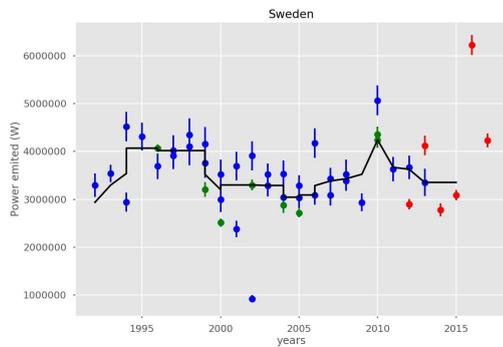 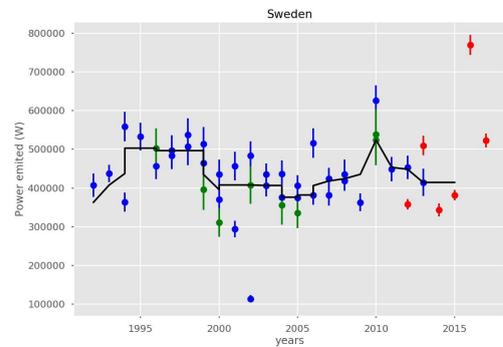

*Fig. B16:* Sunset trends of Sweden.    *Fig. C16:* Midnight trends of Sweden.

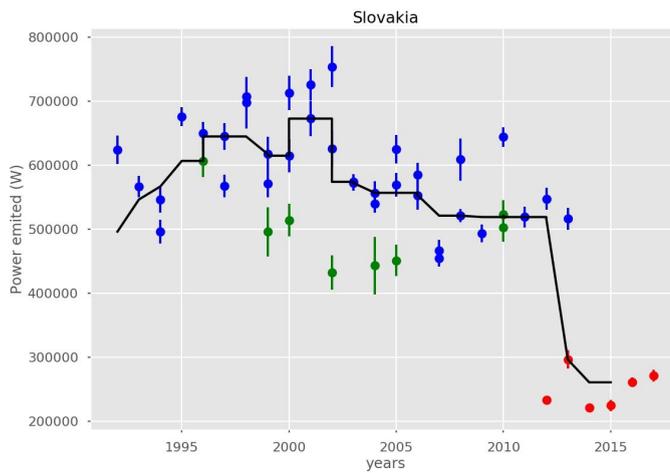

*Fig. A17*: RAW trends of Slovakia. Green data correspond to the calibrated DMSP, blue correspond to non-calibrated DMSP and red correspond to VIIRS. Can be seen that there is a big gap between the DMSP non-calibrated data and the VIIRS. Probably because of the dimming of public lighting.

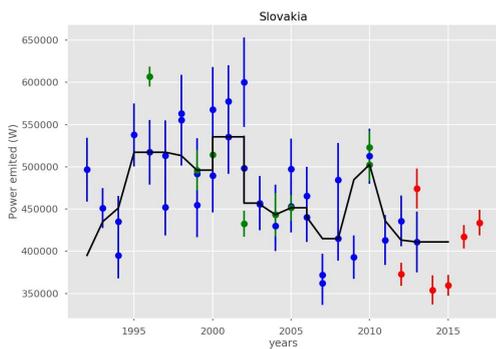

*Fig. B17:* Sunset trends of Slovakia.

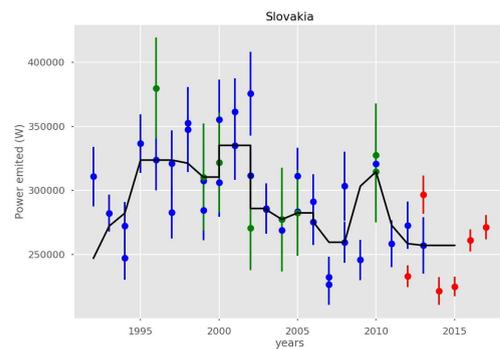

*Fig. C17:* Midnight trends of Slovakia.

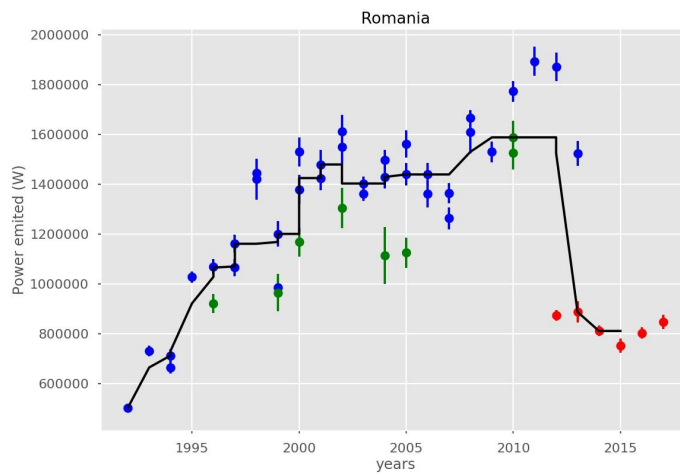

*Fig. A18*: RAW trends of Romania. Green data correspond to the calibrated DMSP, blue correspond to non-calibrated DMSP and red correspond to VIIRS. Can be seen that there is a big gap between the DMSP non-calibrated data and the VIIRS. Probably because of the dimming of public lighting.

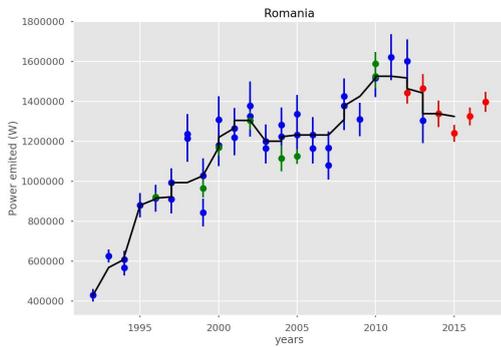

*Fig. B18:* Sunset trends of Romania.

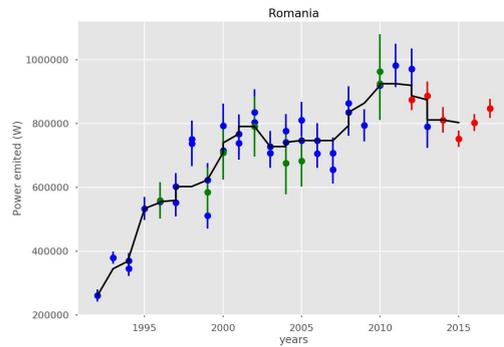

*Fig. C18:* Midnight trends of Romania.

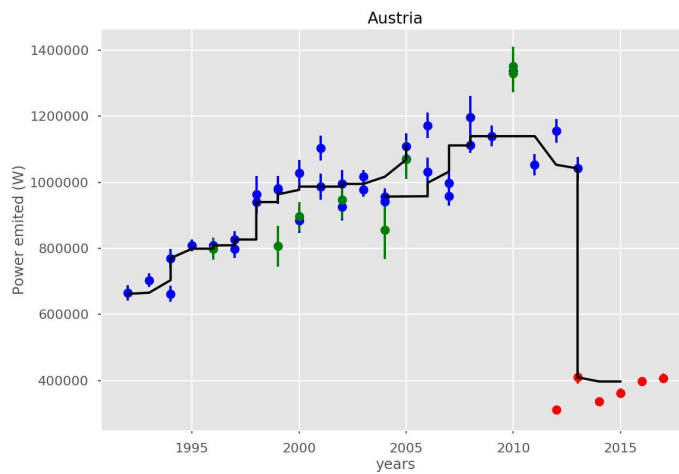

*Fig. A19*: RAW trends of Austria. Green data correspond to the calibrated DMSP, blue correspond to non-calibrated DMSP and red correspond to VIIRS. Can be seen that there is a big gap between the DMSP non-calibrated data and the VIIRS. Probably because of the dimming of public lighting.

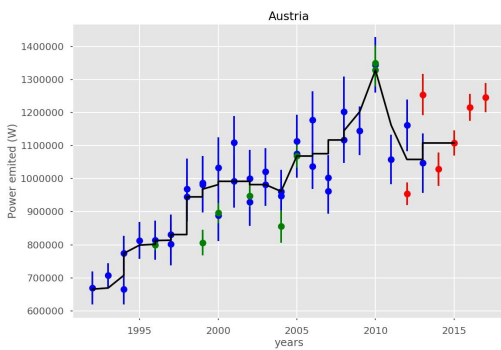

*Fig. B19:* Sunset trends of Austria.

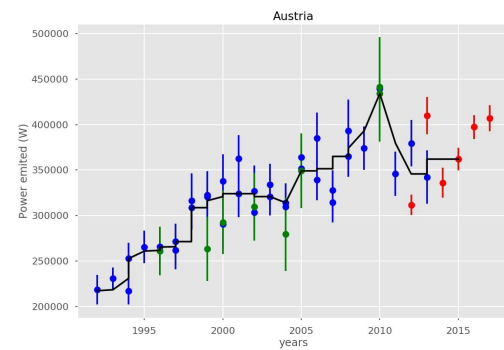

*Fig. C19:* Midnight trends of Austria.